\documentclass{article}
\title{ON SOME RECENT RESULTS ABOUT INERTIAL MANIFOLDS AND KINEMATIC DYNAMOS}
\author{Manuel N\'u\~nez\\Departamento
de An\'alisis Matem\'atico\\ Universidad de
 Valladolid\\ $47005$ Valladolid, Spain\\
e-mail: mnjmhd@am.uva.es} 
\date{}

\begin{document}
\maketitle

\begin{abstract}
The conditions imposed in the paper ['Inertial manifolds and completeness
of eigenmodes for unsteady magnetic dynamos', Physica D {\bf 194} (2004) 297-319]
on the fluid velocity to guarantee the existence of inertial manifolds for the
kinematic dynamo problem are too demanding, in the sense that they imply that all the
solutions tend exponentially to zero. The inertial manifolds are meaningful because they
represent different decay rates, but the classical kinematic dynamos where the
magnetic field is maintained or grows are not covered by this approach,
at least until more refined estimates are found.
\end{abstract}
{\bf PACS} (2003): 02.30.Jr, 52.30.Cv, 52.35.Ra, 91.25.Cw
\par\medskip\noindent
\section{Introduction}
In \cite{LH}, the existence of inertial manifolds for the kinematic dynamo
problem under certain conditions is proved. The result is applied to the
case of time-periodic flows in a spatially periodic box $\Omega$, showing
that any solution may be represented as the sum of an exponentially decreasing
function plus a finite sum of Floquet-like terms: time exponentials times
functions periodic in time. This term corresponds to the solutions within
a finite-dimensional inertial manifold ${\cal M}(t)$. The hypotheses needed
to prove the existence of ${\cal M}(t)$ depend, as usual, on a spectral gap
condition: the eigenvalues $(\mu_{N})$ of the Stokes operator must satisfy
for some $N$
\begin{eqnarray}
\frac{2}{\sqrt{\mu_{N+1}}-\sqrt{\mu_{N}}}+\frac{1}{\sqrt{\mu_{N+1}}}
<\frac{\eta}{3w_{0}},
\end{eqnarray}
where $\eta$ is the magnetic diffusivity and $w_{0}$ is the following norm
on the velocity: assume that both the velocity ${\bf v}$ and its gradient
$\nabla{\bf v}$ are uniformly bounded for all time in $\Omega$, and let
$v_{0}$ and $u_{0}$ be their respective maxima. Then $w_{0}=v_{0}+u_{0}\mu_{1}
^{-1/2}$.\par
Inertial manifolds are often elusive objects in fluid dynamic problems, and
this case is no exception. It is apparent that (1) is rather demanding, given
the extremely small diffusivity occurring in realistic dynamo problems. We will
show that in several cases, including the examples in \cite{LH}, (1) implies
that all the solutions tend exponentially to zero.
In fact the conditions for this to occur are weaker than (1). Hence all the
Floquet exponents have negative real part.\par
The inertial manifolds are still interesting because the decay rate within them
is different from the decay rate transverse to them, but the original object of kinematic
dynamo theory, which was to find velocity fields such that the magnetic field
associated to them was maintained, or better grew exponentially, cannot be
achieved with these examples: finer estimates are needed.\par 
Although the authors restrict themselves to the space periodic case, their methods
seem adaptable with minor modifications upon (1) to other bounday value
problems, such as Dirichlet ones. We will also comment briefly upon this case
in order to illustrate the general situation.
\section{Energy inequalities}
The induction equation satisfied by the magnetic field ${\bf B}$ is
\begin{eqnarray}
\frac{\partial{\bf B}}{\partial t}=\eta \nabla^{2} {\bf B}
-{\bf v}\cdot\nabla{\bf B}+{\bf B}\cdot\nabla{\bf v},
\end{eqnarray}
to which it must be added $\nabla\cdot{\bf B}=\nabla\cdot{\bf v}=0$
and adequate boundary conditions: ${\bf B}$ and ${\bf v}$
periodic, ${\bf B}$ of mean zero in $\Omega$
in the periodic case, ${\bf B}\mid_{\partial\Omega}={\bf 0}$ in the
Dirichlet case. Energy inequalities are obtained by the standard
method of multiplying (2) by ${\bf B}$ and integrating in $\Omega$.
The diffusive term equals
\begin{eqnarray}
\eta\int_{\Omega}\nabla^{2}{\bf B}\cdot{\bf B}\,dV=\eta\int_{\partial\Omega}
{\bf B}\cdot\frac{\partial {\bf B}}{\partial n}\,d\sigma-\eta
\int_{\Omega}|\nabla {\bf B}|^{2}\,dV.
\end{eqnarray}
and the boundary term vanishes for these boundary conditions:
obviously for the Dirichlet case, and in periodic problems because
${\bf B}$ is periodic and the normal vector antiperiodic at opposite
sides of the box. Also the lagrangian term vanishes:
\begin{eqnarray}
\int_{\Omega}({\bf v}\cdot\nabla {\bf B})\cdot{\bf B}\,dV=
\frac{1}{2}\int_{\Omega}{\bf v}\cdot \nabla B^{2}\,dV=
\frac{1}{2}\int_{\partial \Omega}B^{2}{\bf v}\cdot{\bf n}\,d\sigma=0.
\end{eqnarray} 
As for the remaining term, it may be written in two ways. Directly
\begin{eqnarray}
\int_{\Omega}{\bf B}\cdot\nabla{\bf v}\cdot{\bf B}\,dV,
\end{eqnarray}
or, after integration by parts,
\begin{eqnarray}
\int_{\Omega}{\bf B}\cdot\nabla{\bf v}\cdot{\bf B}\,dV=
\int_{\Omega}{\bf B}\cdot\nabla({\bf B}\cdot{\bf v})-
{\bf B}\cdot\nabla{\bf B}\cdot{\bf v}\,dV\nonumber\\
=\int_{\partial\Omega}({\bf B}\cdot{\bf v}){\bf B}\cdot
{\bf n}\,d\sigma-\int_{\Omega}{\bf B}\cdot\nabla{\bf B}
\cdot{\bf v}\,dV,
\end{eqnarray}
and again the boundary integral vanishes. All this is classical
(see e.g. \cite{Te}). Therefore, for any $\alpha\in[0,1]$, we
may write
\begin{eqnarray}
\frac{1}{2}\frac{\partial}{\partial t}\int_{\Omega}B^{2}\,dV=-\eta\int_{\Omega}|\nabla{\bf B}|^{2}
\,dV\nonumber\\+\alpha\int_{\Omega}{\bf B}\cdot\nabla{\bf v}\cdot{\bf B}\,dV-
(1-\alpha)\int_{\Omega}{\bf B}\cdot\nabla{\bf B}\cdot{\bf v}\,dV.
\end{eqnarray}
Denoting by $|\quad|$ the $L^{2}(\Omega)$-norm, and using elementary
bounds,
\begin{eqnarray}
\frac{1}{2}\frac{\partial}{\partial t}|{\bf B}|^{2}\leq -\eta|\nabla{\bf B}|^{2}+\alpha u_{0}
|{\bf B}|^{2}+(1-\alpha) v_{0}|{\bf B}||\nabla{\bf B}|.
\end{eqnarray}
As asserted, denote by $0<\mu_{1}<\mu_{2}<...$ the eigenvalues of the Stokes
operator (which coincide with those of minus the laplacian) in the space
under consideration: $H_{per}^{2}(\Omega)$ for the periodic case,
$H^{2}(\Omega)\cap H_{0}^{1}(\Omega)$ for Dirichlet conditions. Then
\begin{eqnarray}
|\nabla {\bf B}|^{2}=\sum_{j=1}^{\infty}\mu_{j}|{\bf B}_{j}|^{2}
\geq\mu_{1}\sum_{j=1}^{\infty}|{\bf B}_{j}|^{2}=\mu_{1}|{\bf B}|^{2},
\end{eqnarray}
where ${\bf B}_{j}$ is the $j$-th component of the field in the orthogonal
base of eigenvectors of the Stokes operator. Hence
\begin{eqnarray}
\frac{1}{2}\frac{\partial}{\partial t}|{\bf B}|^{2}\leq
-\eta|\nabla {\bf B}|^{2}+\alpha u_{0}\mu_{1}^{-1}|\nabla{\bf B}|^{2}
+(1-\alpha)v_{0}\mu_{1}^{-1/2}|\nabla {\bf B}|^{2}.
\end{eqnarray}
If $-k=-\eta+\alpha u_{0}\mu_{1}^{-1}+(1-\alpha)v_{0}\mu_{1}^{-1/2}<0$,
we have an inequality 
\begin{eqnarray}
\frac{1}{2}\frac{\partial}{\partial t}|{\bf B}|^{2}\leq -k|\nabla
{\bf B}|^{2}\leq -k\mu_{1}|{\bf B}|^{2},
\end{eqnarray}
which implies that any solution decays exponentially in the $L^{2}(\Omega)$-norm.
\section{Analysis of the estimates}
Obviously the condition
\begin{eqnarray}
\alpha u_{0}\mu_{1}^{-1}+(1-\alpha)v_{0}\mu_{1}^{-1/2}<\eta,
\end{eqnarray}
holds for some $\alpha\in[0,1]$ if and only if it holds at some of the
extremes of the interval, i.e.
$u_{0}\mu_{1}^{-1}<\eta$ or $v_{0}\mu_{1}^{-1/2}<\eta$. Notice that
both of these quantities are smaller than $w_{0}\mu_{1}^{-1/2}$.\par
Let us make a small insert to comment that neither the estimates
in \cite{LH} nor the previous ones are modified by scale changes.
This is because if ${\bf B}(t,{\bf x})$ is a solution of the kinematic
dynamo problem in $\Omega$ with velocity ${\bf v}(t,{\bf x})$ and
the boundary conditions, the solution in $R\Omega$ is ${\bf B}
(R^{-2}t, R^{-1}{\bf x})$, associated to the velocity
$R^{-1}{\bf v}(R^{-2}t,R^{-1}{\bf x})$. The eigenvalues of the Stokes
operator become now $R^{-2}\mu_{N}$. Hence the new value of $v_{0}$
is $R^{-1}v_{0}$, the one of $u_{0}\mu_{1}^{-1/2}$ is $R^{-1}
u_{0}\mu_{1}^{-1/2}$, and therefore (1) holds equally. Also the new
values of $v_{0}\mu_{1}^{-1/2}$ and $u_{0}\mu_{1}^{-1}$ coincide with
the previous ones, so that any of the bounds on them holds. Hence
we may restrict ourselves to domains of fixed size when studying these
problems.\par
Let us compare (12) with (1). For Dirichlet problems in general domains, the
classical theorem of Rayleigh-Faber-Krahn \cite{Fa,Kr} states that the domain
with minimal $\mu_{1}$ among those of given measure is given by the ball.
Thus we can restrict ourselves to balls of radius 1, whose first eigenvalues
are given by the squares of the smallest zero of the Bessel functions
in dimension two, or the spherical Bessel functions in dimension three.
Since those are well known, we can assert
\begin{eqnarray}
\mu_{1}^{-1}u_{0},\mu_{1}^{-1/2}v_{0}< \frac{1}{2.4048}w_{0},
\end{eqnarray}
in dimension two, and
\begin{eqnarray}
\mu_{1}^{-1}u_{0},\mu_{1}^{-1/2}v_{0}< \frac{1}{\pi}w_{0},
\end{eqnarray}
in dimension three. Hence any estimate of the form $w_{0}<r\eta$
is improved by $v_{0}\mu_{1}^{-1/2}$ and $u_{0}\mu_{1}^{-1}$.\par
For periodic problems, all the eigenvalues are well known. In
particular, for the case studied in \cite{LH} of square two
and three-dimensional boxes, $\mu_{1}=1$ and $\sqrt{\mu_{N+1}}
-\sqrt{\mu_{N}}\leq 1$. Using this, it is proved in the paper
that an inertial manifold exists in dimension two if $w_{0}<\eta
/6$, in dimension three if $w_{0}<\eta/12$. This obviously implies
that (12) holds even for all $\alpha$, and by a large margin.
Hence all solutions decay exponentially.
Thus the examples do not cover kinematic dynamos with nondecaying
magnetic fields, but this should not detract from
the fact that the argument is correct. The task is to refine
the estimates in (1) so that they are weaker than the conditions for
general decay. Let us mention that a solution bounded in $L^{2}$-norm
is also uniformly bounded: see \cite{Nu}.
\section{Conclusions}
The conditions put forward in \cite{LH} for the existence of finite-dimensional
inertial manifolds for the kinematic dynamo problem in the space periodic
case turn out to be so strong that all the solutions tend exponentially
to zero. A similar situation is likely to occur for other boundary conditions.
Therefore the results cannot be directly applied to classical kinematic
dynamos where the magnetic field is at leat maintained, at least until
refined estimates are found.

\end{document}